%Paper: hep-th/9502097
%From: "F." Krahe <krahe@stp.dias.ie>
%Date: Wed, 15 Feb 1995 20:06:16 +0000 (GMT)

\font\magnifiedtwelverm=cmr12 scaled\magstep3
\font\magnifiedtwelvebf=cmbx12 scaled\magstep3
\font\magnifiedtwelveit=cmti12 scaled\magstep3
\font\magnifiedtwelvesl=cmsl12 scaled\magstep3

\def\Bigtype{\let\rm=\magnifiedtwelverm \let\bf=\magnifiedtwelvebf
\let\it=\magnifiedtwelveit \let\sl=\magnifiedtwelvesl
\rm}

\font\twelverm=cmr12
\font\twelvebf=cmbx12
\font\twelveit=cmti12
\font\twelvesl=cmsl12

\def\bigtype{\let\rm=\twelverm \let\bf=\twelvebf
\let\it=\twelveit \let\sl=\twelvesl
\rm}

\def\d{\partial}
\def\dq{\partial\cdot\partial}

\def\dm{\partial^\mu}

\def\do{\partial^0}
\def\dM{\partial_\mu}
\def\dN{\partial_\nu}

\def\dsm{\mathop{\dm}^{\leftrightarrow}}

\def\dso{\mathop{\do}^{\leftrightarrow}}

\def\={\mathop{=}^{\rm def}}
\def\intx{\int\limits_{t={\rm const.}} d^3\vec x}

\def\Ama{A^\mu_a}   \def\AMa{A_{\mu a}}
\def\Ana{A^\nu_a}   
   
   \def\ANb{A_{\nu b}}

\def\Fmna{F^{\mu\nu}_a}    
    
\def\Fmnc{F^{\mu\nu}_c}    

\def\tu{\tilde{u}}
\def\ua{u_a}  \def\ub{u_b}  
\def\tua{\tu_a}  \def\tub{\tu_b}  \def\tuc{\tu_c}

{\Bigtype {\bf   On the Algebra of Ghost Fields }}
{\bigtype \hphantom {aaaaa} DIAS-STP-95-02}

\vskip 1.5cm

{\bigtype F.Krahe}

\vskip 0.2cm

{\bigtype {\it Institute for Advanced Studies,
10 Burlington Road, IRL-Dublin 4, Ireland }}

\vskip 0.5cm

Work supported by Schweizerischer Nationalfonds

\vskip 1cm

{\bigtype {\bf Summary. - }} We study in detail the algebra of
free ghost fields which we realize in a Hilbert-Fock
space with positive metric. The investigation of causality
clarifies the exact reason for the failure of the spin-statistics
theorem and leads to the introduction of the Krein Operator. We
study the charge algebra of the ghost fields which gives a
representation of ${\rm gl}(2,{\cal C})$. The symmetries of the
$S$-matrix in ghost space are pointed out.

\vskip 2cm

{\bigtype {\bf 1. - Introduction }}

\vskip 0.2 cm

Ghost fields play a vital role in the quantization of non
abelian gauge theories. Usually being derived in the framework
of path integral quantization [1,2], they have been studied using
methods of operator quantization, too. Some authors have
proceeded along the canonical road [3], hoping that lack of mathematical
precision - caused by the very nature of canonical field quantization
- may be overcome by renormalization. There, considerable progress
has been made by exploiting BRS symmetry [4]. Non abelian gauge
theories have also been studied in the framework of axiomatic
quantum field theory [5]. This approach
guarantees mathematical consistency by deriving consequences from
the fundamental axioms, although the construction of interacting
quantum gauge fields (in four dimensions) has not been achieved so
far. The operator quantization of gauge theories is usually carried
out in the framework of indefinite metric state space [6,7].

Beside these important but difficult matters, fortunately, very
simple questions can be asked about ghost fields, too: What is the
exact reason for the failure of the spin-statistics theorem? Is it
possible to quantize ghost fields in a positive metric Hilbert
space? In what sense are the two ghost fields $u$ and $\tu$
conjugate to each other?
One would reasonably expect the answers to these structural
questions to be independent of the particular interactions the ghost
fields are subjected to.
Therefore, this paper gives a detailed investigation of free ghost
fields. In this case the above questions can be unambiguously
answered, and we will show how they are related to the causality
structure of the ghost fields.

The simple case of free ghost fields also deserves consideration for
the following reason: Recently non abelian gauge theory has been
extensively studied in the framework of causal perturbation theory
[8-11]. There [12-15] only free field operators are used to construct
the functional $S$-matrix $S[g]$. Therefore, detailed knowledge of
their properties is quite useful. The (extended) functional
$S$-matrix $S[g,j]$ also allows for the definition of
interacting quantum fields by functional differentiation [6,12-16]
and these share many structural properties with the free ones [16].

The mathematical tools used in this paper are very standard methods
from the apparatus of second quantization [15,17-20]. Our main task
is to tailor this well known material into a form suitable for the
somewhat unconventional ghost fields.

The paper is organized as follows: The next chapter gives the
construction of the ghost fields in a Hilbert-Fock space with
positive metric. The spin-statistics theorem is dealt with in
chapter 3. Chapter 4 studies charge algebra of
the ghost fields as a representation of ${\rm gl}(2,{\cal C})$.
The construction of the Krein operator is given in chapter 5.
The last chapter discusses the ghost charge conjugation and
symmetries in ghost space.

\vfil\eject

{\bigtype {\bf 2. - Ghost Fields in Hilbert-Fock Space}}

\vskip 0.2 cm

In this chapter we give the detailed construction of the free ghost
fields, $u_a(x)$ and $\tu_a(x)$, as used in [8-11]. These are two
operator valued distributions acting in the ghost Hilbert-Fock space
$H_g$ which satisfy the Klein-Gordon equation:
$$ (\d\cdot\d +m^2)u_a(x)=(\d\cdot\d +m^2)\tu_a(x)=0 \eqno (2.1)$$
and the following anticommutation relations:
$$\{u_a(x),\tu_b(y)\}_+={1\over i}\delta_{a,b}D(x-y),\quad
\{u_a(x),u_b(y)\}_+=\{\tu_a(x),\tu_b(y)\}_+=0 \eqno (2.2)$$
$x$ and $y$ are points in Minkowski space ${\cal R}^4$. $D(x)$ is the
Pauli-Jordan commutation function [1].
The above equation
has to be understood in the sense of tempered distributions [6], i.e.
both sides have to be \lq\lq integrated" with tempered test functions
$F(x,y)$. These free ghost fields are important for the causal
construction of non abelian gauge theories [8-11]. As the gauge
fields themselves, they are in the adjoint representation of the
Lie algebra $g$ of the (global) gauge group $G$ [1]. The index $a$ which
runs from $1$ to dim $G$ refers to an arbitrary but fixed set
$\{T^a\}$ of generators in $g$. We assume $g$
to be semi-simple and compact, i.e.
the Cartan metric of $g$ used for $g$-covariant summation is the
Kronecker tensor. For the free fields considered here the group
index $a$ will play a completely trivial role, for it is only the
construction of the interaction [8-11] where $g$ enters via the
structure constants.

In the following we use covariant notation. The mass $m$ may be
positive or zero. Let $\vec p$ be a given 3-vector. We assign to
it a 4-vector $p$ on the mass-hyperboloid (or cone, resp.)
$\Gamma_m$ by
$$p\= (p^0,\vec p), \quad
p^0\= p^0(\vec p,m)\=\sqrt{{\vec p}^{\, 2}+m^2} \eqno (2.3)$$
The invariant volume-measure on $\Gamma_m$  and the corresponding
Dirac distribution are defined by
$$ dp\= {d^3\vec p\over 2p^0(2\pi)^3},\qquad
\delta(p-p')\= 2p^0(2\pi)^3\delta(\vec p-\vec {p'}) \eqno (2.4)$$
The n-particle Hilbert space $H^{(n)}_g$ for the ghost
fields is defined as follows: Its elements are $L^2$-functions of n
momenta, n group indices, and n ghost indices:
$$ H^{(n)}_g\ni\phi^{(n)}=\phi^{(n)}_{a_1,\cdots ,a_n;
i_1,\cdots ,i_n}(p_1,\cdots ,p_n) \eqno (2.5)$$
which are completely antisymmetric under the simultaneous
exchange of arguments and indices: $(p_i,a_i,i_i)\leftrightarrow
(p_j,a_j,i_j)$. $p=p(\vec p,m)$ is the four-momentum specified above,
the $G$-index $a$ runs from
$1$ to dim $G$, and the ghost index $i$ can take the values $\pm
1$, corresponding to a ghost or an antighost particle, respectively.
The scalar product in $H^{(n)}_g$ is defined by
$$(\psi^{(n)},\phi^{(n)}):= \sum_{a_1,\cdots ,a_n=1}^{{\rm dim}G}
\sum_{i_1,\cdots ,i_n=\pm 1} \int dp_1\cdots dp_n
\overline{\psi}_{a_1,\cdots ,a_n;i_1,\cdots ,i_n}(p_1,\cdots ,p_n)
\phi_{a_1,\cdots ,a_n;i_1,\cdots ,i_n}(p_1,\cdots ,p_n) \eqno
(2.6)$$
The elements of $H^{(n)}_g$ have to have finite norm: $\Vert\phi^{(n)}
\Vert=(\phi^{(n)},\phi^{(n)})^{1/2}<\infty$. $H^{(0)}_g={\cal C}$, with
scalar product
$(\psi^{(0)},\phi^{(0)})=\overline{\psi^{(0)}}\phi^{(0)}$.
The Hilbert-Fock space $H_g$ for the ghost fields is the Hilbert space sum
of all n-particle spaces:
$$ H_g=\bigoplus_0^\infty H^{(n)}_g \eqno (2.7)$$
That means, its elements are sequences
$$H_g\ni\phi=\{\phi^{(n)}\}_{n=0}^{\infty}=\{\phi^{(0)},
\phi^{(1)},\cdots,\phi^{(n)},\cdots\},
\qquad \phi^{(n)}\in H^{(n)}_g \eqno (2.8)$$
with scalar product
$$(\psi,\phi)=\sum_{n=0}^{\infty}(\psi^{(n)},\phi^{(n)})
\eqno (2.9)$$
Again, the norm $\Vert\phi\Vert=(\phi,\phi)^{1/2}$ has to be finite.
The vector
$$\phi_0\= \{1,0,0,0,\cdots\} \eqno (2.10)$$
is called the vacuum.

Let $f=\{f(p)\}\in S(\Gamma_m)$ be a tempered test function on
$\Gamma_m$ [6]. Then the
ghost annihilation operators $c_{a;i}(f)$ are defined by
$$\{c_{a;i}(f)\phi\}^{(n)}_{a_1,\cdots,a_n;i_1,\cdots,i_n}
(p_1,\cdots,p_n)=(n+1)^{1/2}\int dp\overline{f(p)}\phi^{(n+1)}
_{a,a_1,\cdots,a_n;i,i_1,\cdots,i_n}(p,p_1,\cdots,p_n) \eqno (2.11)$$
Their adjoints are the ghost creation operators $c^+_{a;i}(f)$ with action
$$ \{c^+_{a;i}(f)\phi\}^{(n)}_{a_1,\cdots,a_n;i_1,\cdots,i_n}
(p_1,\cdots,p_n)=n^{-1/2}[\delta_{a_1,a}\delta_{i_1,i}f(p_1)
\phi^{(n-1)}_{a_2,\cdots,a_n;i_2,\cdots,i_n}(p_2,\cdots,p_n)-$$
$$-\sum_{r=2}^n\delta_{a_r,a}\delta_{i_r,i}f(p_r)\phi^{(n-1)}
_{a_1,\cdots\widehat{a_r}\cdots,a_n;i_1,\cdots\widehat{i_r}\cdots,i_n}
(p_1\cdots\widehat{p_r}\cdots,p_n)] \eqno (2.12)$$
A hat over an index or argument means omitting.
The creation and annihilation operators are bounded operators
defined on the whole of $H_g$ [15]. Their anticommutators are easily
calculated:
$$\{c_{a;i}(f),c^+_{b;j}(g)\}_+=\delta_{a,b}\delta_{i,j}
\int dp\overline{f(p)}g(p)dp,\qquad
\{c_{a;i}(f),c_{b;j}(g)\}_+=\{c^+_{a;i}(f),c^+_{b;j}(g)\}_+=0 \eqno
(2.13)$$
We remark that the formulae (2.11)-(2.13) also hold if $f\in
L^2(\Gamma_m,dp)$.

In the following we will use the distributional operators
$c_{a;i}(p)$ and $c^+_{a;i}(p)$ defined by
$$c_{a;i}(f)=\int dp \overline{f(p)}c_{a;i}(p),\qquad
c^+_{a;i}(f)=\int dp\, c^+_{a;i}(p)f(p) \eqno (2.14)$$
The \lq\lq symbols" $c^{(+)}(p)$ are operator valued distributions, i.e.
denote the (anti)linear continuous mapping
$$ S(\Gamma_m)\ni f\buildrel c^{(+)}(p) \over
\longrightarrow c^{(+)}(f)\in B(H_g) \eqno (2.15)$$
where $B(H_g)$ is the space of the bounded operators over $H_g$.
These distributions are not regular, i.e. there do not exist
$B(H_g)$ valued locally integrable functions $c^{(+)}(p)$ such that the
above equations hold true in the sense of Lebesgue integrals.

However, it is possible to interpret $c^{(+)}(p)$ as functions in the
following sense: We define the dense domain $D_0\subset H_g$ to consist
of all vectors $\phi$ with only finitely many nonvanishing components
$\phi^{(n)}$
which are not merely $L^2$-functions but tempered test functions of
their arguments. Then there are indeed (unbounded) operator valued functions
$c_{a;i}(p)$ which map $D_0$ into itself, defined by
$$[c_{a;i}(p)\phi]^{(n)}_{a_1,\cdots,a_n;i_1,\cdots,i_n}(p_1,\cdots,p_n)=
(n+1)^{1\over 2}\phi^{(n+1)}_{a,a_1,\cdots,a_n;i,i_1,\cdots,i_n}
(p,p_1,\cdots,p_n) \eqno (2.16)$$
Such the first of eq.(2.14) can be interpreted as a Lebesgue integral
in the sense of unbounded operators from $D_0$ into itself. The adjoint
operators $c^+_{a;i}(p)$ are formally defined by
$$[c^+_{a;i}(p)\phi]^{(n)}_{a_1,\cdots,a_n;i_1,\cdots,i_n}
(p_1,\cdots,p_n)=n^{1\over 2}[\delta_{a_1,a}\delta_{i_1,i}\delta(p_1-p)
\phi^{(n-1)}_{a_2,\cdots,a_n;i_2,\cdots,i_n}(p_2,\cdots,p_n)-$$
$$-\sum_{r=2}^n
\delta_{a_r,a}\delta_{i_r,i}\delta(p_r-p)\phi^{(n-1)}_{a_1,\cdots,
\widehat{a_r},\cdots,a_n;i_1,\cdots,\widehat{i_r},\cdots,i_n}
(p_1,\cdots,\widehat{p_r},\cdots,p_n)\eqno (2.17)$$
However, the vector $c^+_{a;i}(p)\phi $ is, for $\phi\neq 0$,
due to the presence of the Dirac distribution an element of $D'_0$, the
topological dual space of $D_0$, but is not in $H_g$, i.e. the adjoint
operators $c^+_{a;i}(p)$ are defined on the zero vector only. Despite
that the functions $c^+_{a;i}(p)$ exist as sesquilinear forms over $D_0\otimes
D_0$ defined by
$$D_0\otimes D_0\ni [\phi,\psi]\rightarrow c^+_{a;i}(p)[\phi,\psi]\=
{}_{D_0}(\,\phi,c^+_{a;i}(p)\psi)_{D'_0}\eqno(2.18)$$
where this \lq\lq scalar product" means the application of the functional
in $D'_0$ to the vector in $D_0$. By construction, $c^+(p)$ is the
sesquilinear form adjoint of $c(p)$:
$$c^+_{a;i}(p)[\phi,\psi]=\overline{c_{a;i}(p)[\psi,\phi]}\=\overline
{(\psi,c_{a;i}(p)\phi)}=(c_{a;i}(p)\phi,\psi) \eqno (2.19)$$
Normal products [13,20] of the $c^{(+)}(p)$ are defined as
sesquilinear forms over $D_0\otimes D_0$ in the same way, f.e.
$$\{c^+_{a;i}(p)c_{b;j}(q)\}[\phi,\psi]\=
(c_{a;i}(p)\phi,c_{b;j}(q)\psi) \eqno (2.20)$$
The distributional form of eq.(2.13)is
$$\{c_{a;i}(p),c^+_{b;j}(p')\}_+=\delta_{a,b}\delta_{i,j}\delta(p-p'),
\qquad \{c_{a;i}(p),c_{b;j}(p')\}_+=
\{c^+_{a;i}(p),c^+_{b;j}(p')\}_+=0 \eqno (2.21)$$

Now we will define the distributional ghost field
operators in coordinate space, $u_{a;i}(x)$. These are linear
combinations of the expressions
$\int dp\, c^{(+)}_{a;i}(p)e^{\mp ipx}$. We want the
anticommutators of the $u_{a;i}(x)$ to be diagonal in the $G$-index
$a$. Since the $c^{(+)}_{a;i}(p)$ have already this property we do
not discuss $G$-mixing. Then the general Ansatz is
$$u_{a;i}(x)=\int
dp\, [A_{i,j}c_{a;j}(p)e^{-ipx}+B_{i,j}c^+_{a;j}(p)e^{ipx}] \eqno
(2.22)$$
Here $A=(A_{i,j})=\left(\matrix{A_{1,1}&A_{1,-1}\cr
A_{-1,1}&A_{-1,-1}\cr}\right)
, B=(B_{i,j})$ are any two 2x2 matrices. Ghost indices $i,j,\cdots$
appearing at least twice in a monomial are understood to be summed over.
It follows from the preceding remarks that (2.22) is to be
understood in the sense of bilinear forms over $D_0\otimes D_0$.
However, let $F=\{F(x)\}\in S({\cal R}^4)$ be a tempered test function over
Minkowski space. Then the smeared bilinear forms
$$u_{a;i}(F)=\int d^4x\,u_{a;i}(x) F(x) \eqno (2.23)$$
are generated by bounded operators. This defines the ghost fields
as operator valued distributions over $S({\cal R}^4)$. An equivalent,
more direct definition of these operators is provided by
$$u_{a;i}(F)=A_{i,j}c_{a;j}(\tilde F)+B_{i,j}c^+_{a;j}(\hat F),\,\,\,
S(\Gamma_m)\ni\hat F(p)\=\int d^4xF(x)\exp\{ipx\},\,\,
\tilde F(p)\=\overline{\hat F(-p)} \eqno (2.24) $$

Since $p$ is on the mass-hyperboloid(-cone) we certainly have
$$(\dq +m^2)u_{a;i}(x)=0 \eqno (2.25)$$
The anticommutators are easily calculated:
$$\{u_{a;i}(x),u_{b;j}(y)\}_+={1\over i}\delta_{a,b}[A_{i,k}B_{j,k}
D^+(x-y)-B_{i,k}A_{j,k}D^-(x-y)] \eqno (2.26)$$
Here
$$D^{\pm}(x)=\pm i\int\, dpe^{\mp ipx} \eqno (2.27)$$
are the positive and negative frequency parts of the Pauli-Jordan
commutation function [15]
$$ D(x)=D^+(x)+D^-(x) \eqno (2.28)$$
which is up to a factor the only linear combination of
$D^+$ and $D^-$ with causal support, i.e. vanishes for $x\cdot x<0$.
Since we do not want to violate causality, (2.26) gives the
following constraint:
$$A\cdot B^{\rm tr}=-B\cdot A^{\rm tr} \eqno (2.29)$$
The anticommutators are then given by
$$\{u_{a;i}(x), u_{b;j}(y)\}_+={1\over i}\delta_{a,b}C_{i,j}D(x-y)
\eqno (2.30)$$
where $C:=A\cdot B^{\rm tr}$ is antisymmetrical due to (2.29). This
equation has, of course, many solutions. We choose
$$A=\left(\matrix{1&0\cr 0&1}\right) ,\quad B=\left(\matrix
{0&-1\cr 1&0}\right)=-C \eqno (2.31)$$
This gives the desired ghost fields:
$$u_a(x):=u_{a;1}(x)=\int dp\,
c_{a;1}(p)e^{-ipx}+c^+_{a;-1}(p)e^{ipx} \eqno (2.32)$$
$$ \tua(x):=u_{a;-1}(x)
=\int dp\,(-)c_{a;-1}(p)e^{-ipx}+c^+_{a;1}(p)e^{ipx} \eqno (2.33)$$
$$\{\ua(x),\tub(y)\}_+={1\over i}\delta_{a,b}D(x-y), \qquad
\{\ua(x),\ub(y)\}_+=\{\tua(x),\tub(y)\}_+=0 \eqno (2.34)$$
which is the standard choice for the free ghost anticommutation rules
found in the literature. Moreover, these are exactly the right
anticommutators needed for gauge invariance [8-11]. So we have
succeeded in realizing the ghost fields in a positive metric
Hilbert-Fock space.

We close this chapter by discussing the representation $\{U(a,\Lambda)\}$
of the proper Poincar\'e group $P_+^{\uparrow}$ in $H_g$.
This is essentially the same as the representation of $P_+^{\uparrow}$
for a free hermitian bosonic scalar field [21]:
For a vector $\phi\in H_g$ [see(2.8)] one defines
$$[U(a,\Lambda)\phi]^{(0)}=\phi^{(0)},\quad
n\geq 1\Rightarrow[U(a,\Lambda) \phi]
^{(n)}_{a_1,\cdots,a_n;i_1,\cdots,i_n}(p_1,\cdots,p_n)=$$
$$=\exp\{i(p_1+\cdots+p_n)\cdot a\}
\phi^{(n)}_{a_1,\cdots,a_n;i_1,\cdots,i_n}
(\Lambda^{-1}p_1,\cdots,\Lambda^{-1}p_n) \eqno (2.35)$$
This $U(a,\Lambda)$ is unitary, obeys the spectrum condition [21],
and fulfils
$$U(a,\Lambda)u_{a;i}(x)U(a,\Lambda)^{-1}=u_{a;i}(\Lambda x+a) \eqno
(2.36)$$
i.e. all $2({\rm dim}\,G)$ ghost fields simply transform independently
as scalar fields.

The vacuum vector $\phi_0$ is $P_+^\uparrow $ invariant and is
determined by this property uniquely up to a phase. We now demand the
matrices $A$ and $B$ in (2.22) to be invertible, such that the creation
and annihilation operators $c_{a;i}^{(+)}$ can be linearly expressed in terms
of the ghost fields $u_{a;i}$ and their adjoints $u_{a;i}^+$. Then the
vacuum is also cyclic with respect to these fields, i.e. any vector in
$H_g$ can be approximated with arbitrary precision by applying a polynom
in the (smeared) ghost fields and their adjoints to the vacuum [21].
The discussion of the discrete symmetries $P$, $T$, and $C$ may be
found in [22].

It will not have escaped the reader's attention that
the construction of the free ghost fields given here is indeed
very similar to the definition of free bosonic scalar fields. The
differences are: Firstly, to get fermionic ghosts, we have used
anticommuting creation and annihilation operators. Secondly,
causality forced us to take an unusual mixture of these operators in
the definition of the local fields. These seemingly minor
differences will entail major consequences, none the less.

\vskip 1cm

{\bigtype {\bf 3. -  The Failure of the Spin-Statistics Theorem}}

\vskip 0.2 cm

The ghost fields we have constructed in the preceding chapter are spin zero
fields and they obey anticommutation relations. This seems to be a
contradiction to the famous spin-statistics theorem stating that quantum
fields with (half)integer spin obey (anti)commutation relations.
To clarify the situation let us
shortly remind ourselves of the precise content of the
spin-statistics theorem [21]. First, it states that if a single
irreducible quantized spinor field $\Phi_\alpha$ has causal (anti)commutators
with itself and with its
adjoint $\Phi^+_\alpha$ and vanishing (anti)commutators with the
other fields of the theory, then it has the \lq\lq right " connection
between spin and statistics, that means it has causal
(anti)commutators for (half)integer spin.
Its second part makes a statement about a theory in which different
irreducible spinor fields $\Phi_{\alpha\, ;a}$ and their
adjoints $\Phi^+_{\alpha\,;a}$ have causal
(anti)commutators among each other ($\alpha$ is the spin-, $a$ the
internal index). Here \lq\lq abnormal" relations between spin and
statistics are possible. Then, however, the theory possesses certain
symmetries. This allows for the construction of transformed fields
with \lq\lq normal" (anti)commutation relations.

Since this theorem is derived in the context of axiomatic quantum
field theory, the term \lq\lq quantum field" has to be understood in
the precise sense given in [21], i.e. all the Wightman axioms hold
true. To check whether the ghost fields fulfil the presumptions of
the above theorem we have, among other things, to insure that the
ghost fields $u$ and $\tu$ have causal anticommutators not only
with themselves but with their adjoint fields $u^+$ and $\tu^+$.
This question being of great importance we will not restrict
ourselves to the special choice (2.31) here. Instead we go back to the
general Ansatz (2.22) only constrained by the causality condition
(2.29).
Then the adjoint fields are given by
$$ u^+_{a;i}(x)=\int dp\, \overline{B_{i,j}}c_{a;j}(p)e^{-ipx}+
\overline{A_{i,j}}c^+_{a;j}(p)e^{ipx} \eqno (3.1)$$
The causal anticommutators for two adjoint fields are easily obtained by
taking the adjoint of (2.22)
$$ \{u^+_{a,i}(x),u^+_{b;j}(y)\}_+=i\delta_{a,b}\overline{C_{i,j}}
D(x-y) \eqno (3.2)$$
More interesting are the anticommutators of the ghost fields with
their adjoints:
$$\{u^+_{a;i}(x),u_{b;j}(y)\}_+={1\over i}\delta_{a,b}
[\overline{B_{i,k}}B_{j,k}D^+(x-y)-\overline{A_{i,k}}A_{j,k}D^-(x-y)]
\eqno (3.3)$$
Causality of this expression is equivalent to
$$\overline{B}\cdot B^{tr}=-\overline{A}\cdot A^{tr} \eqno (3.4)$$
Transposing this gives
$$B\cdot B^+=-A\cdot A^+ \eqno(3.5)$$
Since for nonvanishing $A$ and $B$ the matrices on the two sides of
this equation are of opposite definiteness the only solution is
$A=B=0$. With this choice the ghost fields would be the zero fields.
Excluding this trivial case we conclude:

{\sl The ghost fields have non causal anticommutation relations with
their adjoints. Therefore, they do not fulfil the presumptions of
the spin-statistics theorem. Consequently, they are allowed to escape
its conclusions.}

It is easy to check that this non-causal anticommutators is the
only \lq\lq deficiency" of the ghost fields, i.e. all other Wightman axioms
hold true [21].
We should remark that the demand for causal (anti)commutators of
quantum fields with each other and with their adjoints does not only
constitute one presumption of the spin-statistics theorem, it is a
far more general assumption entering the definition of an
axiomatic quantum field theory [21] from the very beginning. This
means that all the theorems of (standard) axiomatic quantum field theory
can not be applied to the ghost fields without further check. One
would naturally expect inconsistencies for fields with acausal
behaviour, especially for the construction of the causal $S$-matrix
[12-15]. The resolution of this problem will be given in chapter 5.
There it will prove useful to have some general
formulae for transformations of the ghost fields at hand. These
are systematically derived in the next chapter.

\vskip 1cm

{\bigtype {\bf 4. - The Charge Algebra of the Ghost Fields}}

\vskip 0.2 cm

In this chapter we are going to study the charges of the ghost
fields, their algebra, and the transformations they induce in $H_g$.
The charges are expressions bilinear in the fundamental fields.
We consider again an arbitrary complex matrix
$a=(a_{i,j})$ acting in a two dimensional complex
formal ghost-antighost vector space $V_2$.
We assign to it the following charge operator $Q=Q(a)$ acting in the
Hilbert-Fock space $H_g$:
$$ Q(a):=\sum_{r;a;i} c_{a;i}^{+}(f_r)a_{i,j}c_{a,j}(f_r) \eqno (4.1)$$
Here ${f_r}$ is any complete orthonormal basis in $L^2(\Gamma_m,dp)$.
These charges are unbounded operators defined on the common dense
domain $D\subset H_g$ which
consists of all vectors $\phi$ with only finitely many nonvanishing
components $\phi^{(n)}$. We obviously have $D_0\subset D\subset
H_g$. The above sums converge strongly on $D$
and the charge operators map this domain into itself.
So their products are well defined, too. The (smeared) creation and
annihilation operators $c^{(+)}(f)$ and the smeared fields
$u^{(+)}(F),\tu^{(+)}(F)$ also map $D$ into itself, thereby
giving a meaning to products of themselves with the charges $Q$.
The charges can also be expressed as
$$ Q(a)=\int dp\, c^+_i(p)a_{i,j}c_j(p) \eqno (4.2)$$
where this integrals have again to be understood in the sense of
sesquilinear forms over $D_0\otimes D_0$, or in the sense of
integrable distributions [6]. Here, and in the following we have
suppressed the $G$-index $a$.

The set $\{a\}$ of all complex two by two matrices forms the
complex Lie algebra ${\rm gl}\,(2,{\cal C})$. A short calculation shows
$$Q([a,b]_-)=[Q(a),Q(b)]_- \eqno(4.3)$$
That means the linear map $a\rightarrow Q(a)$ gives
a representation of ${\rm gl}\,(2,{\cal C})$.
In $\{a\}$ we have the additional antilinear structure of passing
from a matrix $a$ to its adjoint $a^+$. Also this structure is
preserved by the \lq\lq Quantization" (4.1):
$$Q(a^+)=Q(a)^+ \eqno(4.4)$$
Here we have to add the technical remark that the Hilbert space
adjoint of $Q(a)$: $Q(a)^+_H$, may generally be defined on a domain
somewhat larger than $D$. Then we mean by $Q(a)^+$ its restriction to
$D$.

A hermitian basis (over ${\cal C}$) of ${\rm gl}(2,{\cal C})$ is given by the
four Pauli matrices $\{\sigma_s\}_{s=0}^3$
$$\sigma_0=\left(\matrix{1&0 \cr 0&1 \cr}\right),\,
  \sigma_1=\left(\matrix{0&1 \cr 1&0 \cr}\right),\,
  \sigma_2=\left(\matrix{0&-i\cr i&0 \cr}\right),\,
  \sigma_3=\left(\matrix{1&0 \cr 0&-1\cr}\right) \eqno (4.5)$$
The corresponding essentially selfadjoint basis of the charge algebra $\{Q\}$
is given by: The ghost number
$$N=Q(\sigma_0)=\int dp\,[c_1^+(p)c_1(p)+c_{-1}^+(p)c_{-1}(p)] \eqno
(4.6)$$
the ghost charge
$$Q_g=Q(\sigma_3)=\int dp\,[c_1^+(p)c_1(p)-c_{-1}^+(p)c_{-1}(p)]
\eqno (4.7)$$
and two operators
$$\Gamma=Q(\sigma_1)=\int dp\,[c_1^+(p)c_{-1}(p)+c_{-1}^+(p)c_1(p)]
\eqno (4.8)$$
$$\Omega=Q(\sigma_2)={1\over i}\int
dp\,[c_1^+(p)c_{-1}(p)-c_{-1}^+(p)c_1(p)]
\eqno (4.9)$$
which replace one antighost by a ghost and (with a different relative sign)
vice versa.

We now want to calculate the commutators of these charges with the
ghost creation and annihilation operators. The following general
formulae are easily established:
$$[Q(a),c^+_i(p)]_-=c^+_j(p)a_{j,i},\quad
[Q(a),c_i(p)]_-=-a_{i,j}c_j(p),\quad Q(a)\phi_0 =0 \eqno (4.10)$$
This implies the action of these charges on a general state $\phi\in
D$ (We again suppress all $G$-indices):
$$\phi=\{\phi^{(0)},\phi^{(1)}_{i_1}(p_1),
\cdots,\phi^{(n)}_{i_1,\cdots,i_n}(p_1,\cdots,p_n),\cdots\},$$
$$Q(a)\phi=\tilde{\phi}=\{\tilde{\phi}^{(0)},\tilde{\phi}^{(1)}_{i_1}(p_1),
\cdots,\tilde{\phi}^{(n)}_{i_1,\cdots,i_n}(p_1,\cdots,p_n)\cdots\},$$
$$\tilde{\phi}^{(0)}=0,\quad n\ge 1\Rightarrow
\tilde{\phi}^{(n)}_{i_1,\cdots,i_n}(p_1,\cdots,p_n)=
\sum_{r=1}^{n}\left\{\sum_{j_r=\pm 1}
a_{i_r,j_r}\phi^{(n)}_{i_1,\cdots,j_r,\cdots,i_n}(p_1,\cdots,p_n)\right\}
\eqno (4.11)$$
This means that the representation of ${\rm gl}(2,{\cal
C})$ in the $n$ particle space $H^{(n)}_g$ is, for fixed $p$,
the $n$-fold direct sum of the self-representation. It follows that
the charge algebra $\{Q\}$ is a faithful representation of ${\rm
gl}(2,{\cal C})$.

By inserting the Pauli matrices for $a$ in (4.10) we get the
following commutators of our four basis operators in $\{Q\}$ with
the ghost creation and annihilation operators:
$$[N,c_j^+(p)]_-=c_j^+(p),\quad [N,c_j(p)]_-=-c_j(p) \eqno (4.12)$$
$$[Q_g,c_j^+(p)]_-=jc_j^+(p),\quad [Q_g,c_j(p)]_-=-jc_j(p) \eqno
(4.13)$$
$$[\Gamma,c_j^+(p)]_-=c_{-j}^+(p),\quad
  [\Gamma,c_j(p)]_-=-c_{-j}(p) \eqno (4.14)$$
$$[\Omega,c_j^+(p)]_-=ijc_{-j}^+(p) ,\quad
  [\Omega,c_j(p)]_-=ijc_{-j}(p) \eqno (4.15)$$
and their commutators with the local fields are
$$[N,u(x)]_-=-\tu^+(x),\quad [N,\tu(x)]_-=u^+(x) \eqno (4.16)$$
$$[Q_g,u(x)]_-=-u(x),\quad [Q_g,\tu(x)]_-=\tu(x) \eqno (4.17)$$
$$[\Gamma,u(x)]_-=\tu(x),\quad [\Gamma,\tu(x)]_-=u(x) \eqno (4.18)$$
$$[\Omega,u(x)]_-=-i\tu(x),\quad [\Omega,\tu(x)]_-=iu(x) \eqno (4.19)$$

It is remarkable that the operator $N$ is distinguished from the other
three charges $Q_g,\Gamma,\Omega$ for two very different reasons.
Firstly, from the algebraic point of view: Its linear span $\lambda N,
\lambda\in{\cal C},$ is the center of the charge algebra $\{Q\}$
(as the linear span of $\sigma_0$ is the center of ${\rm gl}(2,{\cal C})$).
Secondly, form the standpoint of causal field theory: Its commutators with
the local fields $u(x),\tu(x)$ give the not relatively local
fields [21] $u^+(x),\tu^+(x)$, while the commutators of the other
charges with the local fields give back local fields.

This is further illustrated by writing down the following conserved currents:
$$j^\mu_N(x)=i:u^+(x)\dsm u(x):,\quad j^\mu_g(x)=i:\tu(x)\dsm u(x):$$
$$j^\mu_u(x)=i:u(x)\dsm u(x):,\quad
j^\mu_{\tu}(x)=i:\tu(x)\dsm\tu(x): \eqno (4.20)$$
Here the double dot means normal ordering [2,6,15,20]. These
Wick powers [6,20] of the free ghost fields are again operator
valued distributions over $S({\cal R}^4)$, i.e. they have to be
smeared with tempered test functions $F(x)$ over Minkowski space.
They are (after smearing) unbounded operators defined on the dense
domain $D_0$ (see chapter 2) and map this domain into itself.
$D_0$ is actually a dense, common, and invariant domain
for all operators over $H_g$ appearing in the paper at hand. The currents can,
again, likewise be interpreted as functions with values in the
sesquilinear forms over $D_0\otimes D_0$.

The charges are related to these currents by
$$N=\intx j^0_N(x),\quad Q_g=\intx j^0_g(x)$$
$$\Gamma={1\over 2}[Q_u-Q_{\tu}]\={1\over 2}\left[\,\intx j^0_u(x)-
\intx j^0_{\tu}(x)\right]$$
$$\Omega={i\over 2}[Q_u+Q_{\tu}]\=
{i\over 2}\left[\,\intx j^0_u(x)+\intx j^0_{\tu}(x)\right] \eqno
(4.21)$$
These formulae have to be understood in the sense of bilinear forms
over $D_0\otimes D_0$, or in the sense of integrable distributions
[6].

The current $j_N$ uses $u^+$, so it is not a relatively local quantum
field. No wonder then that the derivation implemented by its charge
$N$ acts nonlocally on the field algebra. The three other currents,
however, are relatively local to the basic fields $u,\tu$. The
existence of three bilinear, conserved, local currents for a
scalar field is a unique feature of the ghost field (The more
conventional bosonic charged scalar field has only one such
current).

We now pass from the Lie algebra ${\rm gl}(2,{\cal C})$ to the
corresponding Lie group ${\rm GL}(2,{\cal C})$, i.e. the set
$\{A\}$ of all complex two by two matrices $A$ with ${\rm
det}A\neq 0$, the group product being the usual matrix multiplication.
Again, we look at it as a $4$-dimensional complex Lie group. Each
group element can be written in the form
$$A=\exp\{ia\}=\exp\{iz_s\sigma_s\},\quad z_s=x_s+iy_s\in[-\pi,\pi)+i{\cal R}
\eqno (4.22)$$
In $\{A\}$ we also have the anti-isomorphism of passing from a matrix
$A$ to its adjoint $A^+$. Real $z$'s represent unitary matrices,
obviously.

We now assign to each $A$ the following transformation operator
$T$=$T(A)$ in $H_g$:
$$T(A)=T(\exp\{ia\})\=\exp\{iQ(a)\} \eqno (4.23)$$
Like the charges $\{Q\}$ the transformations $\{T\}$ are defined on
the domain $D$ and map it into itself. The exponential on the right
side of the last equation is defined by its power series which
converges strongly on $D$. Using the Baker-Campbell-Hausdorff
formula and (4.3) gives
$$T(AB)=T(A)T(B) \eqno (4.24)$$
So the transformations $\{T\}$ give a representation of ${\rm
GL}(2,{\cal C})$. Inserting (4.4) into (4.24) gives in addition
$$T(A^+)=T(A)^+ \eqno (4.25)$$
i.e. the \lq\lq quantization" preserves the anti-isomorphism. For the
definition of $T^+$ the technical remark after (4.4) applies.
The quantization of unitary matrices are isometries on $D$. These
can, of course, be extended to unitary operators on $H_g$, and this
process is always understood to have been carried out.

We now study the adjoint action of the transformations $\{T\}$
in $H_g$. The following formulae hold true on $D$:
$${\rm Ad}T(A)[c^+_i(f)]\= T(A)c^+_i(f)T^{-1}(A)=c^+_j(f)A_{j,i}
\eqno (4.26)$$
$${\rm Ad}T(A)[c_i(f)]\= T(A)c_i(f)T^{-1}(A)= A^{-1}_{i,j}c_j(f)
\eqno (4.27)$$
This implies the following action of the transformations
$\{T\}$ on the states $\phi\in D$:
$$\phi=\{\phi^{(0)},\phi^{(1)}_{i_1}(p_1),\cdots,
\phi^{(n)}_{i_1,\cdots,i_n}(p_1,\cdots,p_n),\cdots \}$$
$$T(A)\phi=\phi'=\{\phi'^{(0)},\phi'^{(1)}_{i_1}(p_1),\cdots,
\phi'^{(n)}_{i_1,\cdots,i_n}(p_1,\cdots,p_n),\cdots \}$$
$$\phi'{(0)}=\phi^{(0)},\quad n\geq 1\Rightarrow
\phi'^{(n)}_{i_1,\cdots,i_n}(p_1,\cdots,p_n)=
\sum_{j_1,\cdots,j_n=\pm 1}\left\{\left(\prod_{r=1}^n
a_{i_r,j_r}\right)\phi^{(n)}_{j_1,\cdots,j_n}(p_,\cdots,p_n)\right\}
\eqno (4.28)$$
This means that the representation of ${\rm GL}(2,{\cal C})$ in the
$n$ particle space $H^{(n)}_g$ is, for fixed $p$, the $n$-fold
direct product of the self-representation. It follows that the
transformation group $\{T\}$ is a faithful representation of ${\rm
GL}(2,{\cal C})$.

The adjoint action of the transformation group $\{T\}$ on the charge
algebra $\{Q\}$ is easily calculated by means of (4.26),(4.27):
$${\rm Ad}T(A)[Q(b)]\= T(A)Q(b)T^{-1}(A)=
Q(AbA^{-1})\= Q\left({\rm Ad}A[b]\right) \eqno (4.29)$$
i.e. the adjoint action commutes with the \lq\lq quantization". This will
turn out to be quite useful in the next chapter since it
essentially reduces operator calculations in $H_g$ to matrix algebra
in $V_2$.

We close this algebraic chapter by giving explicitly the adjoint
action of the one parameter groups generated by the charges
$Q(\sigma_s)$ on the creation and annihilation operators and on the
local fields:
$$\eqalignno{
{\rm Ad}\exp\{izN\}[c^+_j(p)]=&\exp\{iz\}c^+_j(p)& (4.30)\cr
{\rm Ad}\exp\{izN\}[c_j(p)]=&\exp\{-iz\}c_j(p)& (4.31)\cr
{\rm Ad}\exp\{izN\}[u(x)]=&(\cos z)u(x) +(\sin z)(i\tu)^+(x)& (4.32)\cr
{\rm Ad}\exp\{izN\}[i\tu(x)]=&-(\sin z)u^+(x)+ (\cos z)i\tu(x)& (4.33)\cr
{\rm Ad}\exp\{izQ_g\}[c^+_j(p)]=&\exp\{jiz\}c^+_j(p)& (4.34)\cr
{\rm Ad}\exp\{izQ_g\}[c_j(p)]=&\exp\{-jiz\}c_j(p)& (4.35)\cr
{\rm Ad}\exp\{izQ_g\}[u(x)]=&\exp\{-iz\}u(x)& (4.36)\cr
{\rm Ad}\exp\{izQ_g\}[\tu(x)]=&\exp\{iz\}\tu(x)& (4.37)\cr
{\rm Ad}\exp\{iz\Gamma\}[c^+_j(p)]=&(\cos z)c^+_j(p)+
i(\sin z)c^+_{-j}(p)& (4.38)\cr
{\rm Ad}\exp\{iz\Gamma\}[c_j(p)]=&(\cos z)c_j(p)-i(\sin z)c_{-j}(p)& (4.39)\cr
{\rm Ad}\exp\{iz\Gamma\}[u(x)]=&(\cos z) u(x)+(\sin z)i\tu(x)& (4.40)\cr
{\rm Ad}\exp\{iz\Gamma\}[i\tu(x)]=&-(\sin z)u(x)+(\cos z)i\tu(x)& (4.41)\cr
{\rm Ad}\exp\{iz\Omega\}[c^+_j(p)]=&(\cos z)c^+_j(p)-
j(\sin z)c^+_{-j}(p)& (4.42)\cr
{\rm Ad}\exp\{iz\Omega\}[c_j(p)]=&(\cos z) c_j(p)-j(\sin z)c_{-j}(p)& (4.43)\cr
{\rm Ad}\exp\{iz\Omega\}[u(x)]=&(\cos z)u(x)+(\sin z)\tu(x)& (4.44)\cr
{\rm Ad}\exp\{iz\Omega\}[\tu(x)]=&-(\sin z)u(x)+(\cos z)\tu(x)& (4.45)\cr
}$$

\vskip 1cm

{\bigtype {\bf 5.- The Construction of the Krein operator}}

\vskip 0.2 cm

We have seen in chapter 3 that the adjoint ghost fields $u^+,\tu^+$
have non causal anticommutators with the ghost fields $u,\tu$.
Since causality is a cornerstone of relativistic quantum
field theory we expect trouble. This is avoided in the most simple
way conceivable: The theory has to be constructed by using only the
local fields $u,\tu$ while the adjoint fields $u^+,\tu^+$ will not
appear at all. In canonical field theory, for example, the
Lagrangian or Hamiltonian has to be (and is!) constructed by solely
using $u$ and$\tu$ (beside other non-ghost fields). An axiomatic
framework, too, would have to exclude $u^+$ and $\tu^+$ from
its field content from the very beginning.

The situation is quite
transparent in causal perturbation theory [12-15] which aims at the
construction of the functional $S$-matrix in the form
of the following power series
$$ S[g]=1+\sum_{n=1}^\infty e^n\int d^4x_1\cdots
d^4x_n\,T^{(n)}(x_1,\cdots,x_n) g(x_1)\cdots g(x_n) \eqno (5.1)$$
Indeed, if only we are given $T^{(1)}(x)$, all higher $T^{(n)}(x)$ can
be derived by merely using causality and Poincar\'e invariance of
$S[g]$ [12-15]! Roughly speaking, $T^{(n)}$ is just the $n$-fold time
ordered product of $n$ times $T^{(1)}$. $T^{(1)}(x)$ is a certain
combination of Wick powers of free quantum fields and its field
content and special form determine the framework of the theory and
its interaction.
Only those fields will appear in $T^{(n)}$ which are already
present in $T^{(1)}$. Therefore, all we have to do is to avoid
using the non causal fields $u^+,\tu^+$ in the construction of $T^{(1)}$
and we will forever have got rid of them.

In massless Yang-Mills theories [8-11], for example,
one considers the interaction
$$T^{(1)}(x)=-{i\over 2}f_{a,b,c}\left\{:\AMa\ANb\Fmnc:(x)
+:\AMa\ub\dm\tuc:(x)\right\}\eqno (5.2)$$
$f_{a,b,c}$ are the structure constants of $G$, $\Ama$ are the free
quantized gauge potentials, $\Fmna$ are the corresponding free field
strengths, and $\ua$ and $\tua$ are our now familiar ghost fields. This
$T^{(1)}(x)$ is invariant under gauge transformations [8] generated by
the differential operator [23]
$${\cal Q}= \intx [(\dN\Ana)\dso\ua](x) \eqno (5.3)$$
Since the noncausal fields $u^+,\tu^+$ do indeed not appear
causality is preserved.

The story, however, goes on. For, by suspending the use of the adjoint
fields we run into the the next problem: The $S$-matrix (5.1) should
be unitary. Then $T^{(1)}(x)$ must be antihermitian. This is,
however, not the case, since the construction of an antihermitian
quantity certainly requires using both the ghost fields and their
adjoints. In canonical quantization it is the Lagrangian or
Hamiltonian which has to be hermitian, and in the axiomatic approach
[21] the adjoint fields always appear on equal footing with the
field themselves.

The solution of this problem is typical to gauge theories. The
$S$-matrix, or $iT^{(1}(x)$ [the Lagrangian, Hamiltonian],
is no longer supposed to be unitary but pseudo-unitary,
or pseudo-hermitian, respectively. This means the following: In the
Hilbert space $H$ of the underlying gauge theory there exists
a distinguished bounded linear operator $J:\, H\rightarrow H$
satisfying
$$J^+=J,\quad J^2=1 \eqno (5.4)$$
i.e. $J$ is hermitian and unitary. It is, however, not a positive
but an indefinite operator.
The pseudo-adjoint $O^K$ of an operator $O$ over $H$ is
then defined by
$$O^K\=JO^+J \eqno (5.5)$$
The rigorous discussion of domain questions can be found in [7].
This pseudo-adjugation shares all the algebraic properties of the
usual adjugation, i.e
$$(O_1+O_2)^K=O_1^K+O_2^K,\,\,(zO)^K=\overline zO,\,\,
(O_1O_2)^K=O_2^KO_1^K,\,\,\left(O^K\right)^K=O \eqno (5.6)$$
Positivity, however, is lost:
$$ \exists O:\,\, O^KO\not\geq 0 \eqno (5.7)$$
An operator $H$ which satisfies $H^K=H$ is called pseudo-hermitian,
an operator $U$ obeying $UU^K=1$ is called pseudo-unitary. For
referring explicitly to the operator $J$ the terms $J$-hermitian and
$J$-unitary are used, too. The operator $J$ is called the
Krein operator and the pair $\{H,K\}$ is
called a Krein space. Let $(\phi,\psi)$ be the (positive definite)
scalar product in $H$. The operator $J$ defines a second, indefinite
scalar product by
$$\langle\phi,\psi\rangle\= (\phi,J\psi) \eqno (5.8)$$
Krein spaces are well studied in the mathematical literature [7] and
they are the appropriate spaces to study quantized gauge theories
[3,5,6,24]. The $J$-unitarity of the $S$-matrix in quantized gauge
theories is important because, together with gauge invariance,
it implies the unitary of the $S$-matrix on the physical subspace
[3,11,15].

Let us go back to the Yang-Mills theory (5.2). Its Hilbert space is
the direct product of the  gauge field Fock space and the ghost Fock
space, and the Krein operator factorizes accordingly:

$$ H=H_A\otimes H_g,\,\,\, J=J_A\otimes J_g \eqno (5.9)$$
$J_A$ is given by
$$J_A\=\prod_{a=1}^{{\rm dim}\,G}(-1)^{N^0_a} \eqno (5.10)$$
where $N^0_a$ is the number operator for ${\rm SO}(3,{\cal
R})$-scalar gauge particles of $G$-colour $a$. The gauge
potentials are pseudo-hermitian with respect to $J_A$:
$$\left(\Ama\right)^K\=J_A\left(\Ama\right)^+J_A=\Ama \eqno (5.11)$$
The reader can find the details of this construction in refs.[11,15].

We, instead, proceed in our discussion of the ghost field algebra.
Our aim is the construction of the Krein operator $J_g$ for the
ghost space $H_g$. This will then define the pseudo-adjoint ghost
fields
$$u^K(x)\=J_gu^+(x)J_g,\,\,\, \tu^K(x)\=J_g\tu^+(x)J_g \eqno
(5.12)$$
The key to the construction of $J_g$ is causality. For, we know that
the $J_g$-adjoint fields $u^K, \tu^K$ have to appear in the
construction of $J_g$-hermitian quantities like $T^{(1)}(x)$.
Since we want to preserve causality these fields have to be
relatively local to the other fields in $T^{(1)}$, especially to the
ghost fields themselves. We know already that the adjoint fields
$u^+,\tu^+$ do not have this property. Therefore, the last equation
tells us that the operator $J_g$ has to restore causality.

This story can be compared to the construction of the
covariant derivative in differential geometry. There the partial
derivative fails to be covariant. Covariance is restored by adding
to it a connection term which takes the parallel transport into
account. Obviously, this can only work because the connection term
is not covariant itself.

By analogy we find that the operator $J_g$ cannot be
(quasi)local itself, i.e. it has to act nonlocally on the ghost
fields. We will construct $J_g$ using the charges and
transformations discussed in the preceding chapter. It follows that
we would not succeed if solely using the quasilocal charges
$Q_g,\Gamma,\Omega$. Instead, we will certainly need the non
quasilocal ghost number $N$.

Since $J_g^2=1$ the first guess would
be to take for $J_g$ the operator
$$E\=(-1)^N=\exp\{i\pi N\} \eqno(5.13)$$
However, $N$ is not the only charge operator with integer spectrum.
The mutually commuting operators
$$N_j\= {1\over 2}[N+jQ_g],\,\,j\in\{1,-1\} \eqno (5.14)$$
which separately measure the number of ghost and antighost particles
have integer spectrum, too. This implies
$$ E=(-1)^N=(-1)^{N_1+N_{-1}}=(-1)^{N_1}(-1)^{N_{-1}}=(-1)^{N_1}
(-1)^{-N_{-1}}=(-1)^{N_1-N_{-1}}=(-1)^{Q_g} \eqno (5.15)$$
showing that this particular function of the not
quasilocal operator $N$ is actually equal to a transformation
generated by the quasilocal operator $Q_g$. Thus it cannot be the right
choice for $J_g$. Instead, $E$ is the grading operator for the
${\cal Z}_2$-graded operator algebra [23] $\{O\}$.
For, if $O_b$ and $O_f$ denote Bose and
Fermi operators, respectively [i.e. even resp. odd polynomials in the
operators $c^{(+)}(f)$], we have
$$EO_bE=+O_b,\,\,E_bO_fE=-O_f \eqno (5.16)$$

We now use one of the operators $N_j$ for the
construction of the Krein operator since these operators still have
an admixture of the nonlocal operator $N$. So we define
$$I\=(-1)^{N_{-1}}=\exp\{i{\pi\over 2}[N-Q]\}=T(\Sigma_3) \eqno
(5.17)$$
We have used a capital letter for the Pauli matrix $\Sigma_3$
because we here interpret it as an element of the
Lie group ${\rm GL}(2,{\cal C})$
while we use small letters if we interpret it as an element of
the Lie algebra ${\rm gl}(2,{\cal C})$. $I$ is indeed
hermitian and unitary. Its action on operators follows from
$$Ic^{(+)}_j(p)I=jc^{(+)}_j(p) \eqno (5.18)$$
Let us denote the pseudo-adjoint with respect to $I$ by a star:
$$O^*\=I0^+I \eqno (5.19)$$
Then we have
$$u^*(x)=\tu(x),\qquad\tu^*(x)=u(x) \eqno (5.20)$$
So the $I$-adjoint of the local fields are local fields again.
This is exactly what we wanted. Moreover, $u$ and $\tu$ are indeed
$I$-adjoint to each other, which answers the last of the three
questions in the introduction. So one might well use the symbol
$\sim$ instead of $*$ for the $I$-adjugation. $I$ would be a perfectly
right choice for the Krein operator. The reason why it is not used,
however, is that neither $iT^{(1)}(x)$ in (5.2) nor ${\cal Q}$ in
(5.3) would be $I$-hermitian. While this would be easily corrected
by giving different though equivalent definitions in (5.2, 5.3),
expressions (5.2, 5.3) have historically preceded the discussion of
the Krein structure of this theory.

The Krein operator $J_g$ is obtained from $I$ by the following
unitary transformation $S$ in $H_g$:
$$ S=T(U),\,\,U\={i\over\sqrt 2}\left(\Sigma_1+\Sigma_3\right) \eqno
(5.21)$$
Since
$$U=\exp\left\{i{\pi\over 2\sqrt 2}\left(\sigma_1+\sigma_3\right)\right\}
\eqno (5.22)$$
we explicitly have
$$ S=\exp\left\{i{\pi\over 2\sqrt 2}\left(\Gamma+Q_g\right)\right\}
\eqno(5.23)$$
The Krein operator is finally defined by
$$J_g\=SIS^{-1} \eqno (5.24)$$
By using the equations obtained in the preceding chapter we find
$$J_g=T\left(\Sigma_1\right)=\exp\left\{i{\pi\over
2}\left(N-\Gamma \right)\right\}=i^{N-\Gamma} \eqno (5.25)$$
Its action on operators follows from
$$J_gc^{(+)}_j(p)J_g=c^{(+)}_{-j}(p) \eqno (5.26)$$
This implies that the $J_g$-adjoint fields in (5.12) are given by
$$ u^K(x)=u(x),\qquad \tu^K(x)=-\tu(x) \eqno (5.27)$$
It is then easy to check that the desired pseudo-hermiticity properties
hold true:

$$\left[T^{(1)}(x)\right]^K=-T^{(1)}(x),\quad {\cal Q}^K={\cal Q}
\eqno (5.28)$$

To obtain these important properties, equations (5.27) have
often been taken as the definition of the pseudo-adjugation $K$
[1,3,11]. Our aim here was to show how this fits naturally
in the framework of Krein spaces, to construct $J_g$ explicitly, and
to show its relation to the causality structure of the ghost fields.

We close this chapter by discussing the $J_g$ adjugation
on the charge algebra $\{Q\}$. This is given by
$$N^K=N,\quad
Q_g^K=-Q_g,\quad\Gamma^K=\Gamma,\quad\Omega^K=-\Omega\eqno (5.29)$$
The set $\{H_J\}\subset\{Q\}$ of all $J_g$-hermitian charges $H_J$,
i.e. the charges satisfying $H_J^K=H_J$,
is consequently given by the real linear span of
$\{N,iQ_g,\Gamma,i\Omega\}$ and forms a real four-dimensional Lie
subalgebra of the complex four dimensional Lie algebra $\{Q\}$.
It follows from the representation properties discussed in the last
chapter that its elements can be written as
$$H_J=Q(m),\quad {\rm with}\,\,\, m=m^k\=\Sigma_1m^+\Sigma_1 \eqno (5.30)$$
The set of the 2x2 matrices $\{m\}$ fulfilling the last equation
forms the real Lie algebra $u(1,1)$. $\{H_J\}$ is therefore a
faithful representation of this Lie algebra.

The $J_g$-unitary transformations $S_J\in\{T\}$, i.e. the
transformations satisfying $S_J^KS_J=1$, form a real four-dimensional
Lie subgroup of the complex four-dimensional Lie group $\{T\}$.
They can be written as
$$S_J=T(M),\quad {\rm with}\,\,\, MM^k\=M\Sigma_1M^+\Sigma _1 =1
\eqno (5.31)$$
The set of the 2x2 matrices $M$ fulfilling the last equation forms
the real Lie group $U(1,1)$. $\{S_J\}$ is therefore a faithful
representation of this Lie group. Its identity component is given by
the exponentiation of $\{H_J\}$.

\vskip 1cm

{\bigtype {\bf 6. - Ghost Charge Conjugation and Symmetries in Ghost
Space }}

\vskip 0.2cm

In this chapter we will discuss the ghost charge conjugation $C_g$.
As the name suggests this is an operator reflecting the ghost
charge, i.e. satisfying
$$C_gQ_gC^{-1}_g=-C_g \eqno (6.1)$$
It is easy to check that the Krein operator $J_g=i^{\Gamma-N}$ would
do this job. To be a symmetry of the theory (of the $S$-matrix, f.e.)
$C_g$ should be a quasilocal operator, however. Such it is easily
constructed by taking the \lq\lq quasilocal part" of $J_g$ ,i.e. we define
$$C_g\= i^NJ_g=i^\Gamma=T(i\Sigma_1) \eqno (6.2)$$
Indeed, since $N$ acts trivially on the charge algebra we have
$$ C_gQ(a)C_g^{-1}=J_gQ(a)J_g,\quad\forall a \eqno (6.3)$$
This gives the action of $C$ on $\{Q\}$ as
$$ C_gNC_g^{-1}=N,\quad C_gQ_gC_g^{-1}=-Q_g,\quad C_g\Gamma C_g^{-1}=\Gamma,
\quad C_g\Omega C_g^{-1}=-\Omega \eqno (6.4)$$
The square of $C_g$ is given by the grading operator (5.15):
$$C_g^2=(i^NJ_g)^2=(-1)^NJ_g^2=(-1)^N=(-1)^Q=E;\quad C_g^4=1 \eqno (6.5)$$
The adjoint of $C_g$ is given by
$$C_g^+=\left[i^NJ_g\right]^+=J_g(-i)^N=J_g^{-1}\left[i^N\right]^{-1}
=C_g^{-1} \eqno (6.6)$$
i.e. $C_g$ is unitary. Since it commutes with $J_g$:
$$ [C_g,J_g]_-=0 \eqno (6.7)$$
it is $J_g$-unitary as well:
$$C_gC_g^K=1 \eqno(6.8)$$
The action of the ghost charge conjugation on the particle operators
follows from (4.38,4.39):
$$C_gc^+_j(p)C_g^{-1}=ic^+_{-j}(p),\quad C_gc_j(p)C_g^{-1}=-ic_{-j}(p)
\eqno (6.9)$$
while (4.40,4.41) imply its action on the local fields:
$$C_gu(x)C_g^{-1}=i\tu(x),\quad C_g\tu(x)C_g^{-1}=iu(x)\eqno
(6.10)$$
This transformation of the ghost fields has also been discussed in
[3].

We remind the reader not to confuse the ghost charge conjugation
$C_g$ with the charge conjugation $C$ which is discussed in [22].
The latter one is the baryonic charge conjugation: The Yang-Mills
fields (gluons) have vanishing baryonic charge, of course. However,
due to their coupling to the baryons (quarks) they have to transform
nontrivially under the conjugation of the baryonic charge.
The ghost fields, in turn, couple to the Yang-Mills fields and
consequently transform nontrivially under $C$ themselves.

Let us go back to the interaction $T^{(1)}(x)$ in eq. (5.2). Beside
being gauge invariant it is invariant under the transformations
generated by the ghost charge:
$$[Q_g,T^{(1)}(x)]_-=0 \eqno (6.11)$$
However, neither the other generators of the transformation group
$\{T\}$: $N$, $\Gamma$, and $\Omega$, nor the ghost charge
conjugation $C_g$ commute with $T^{(1)}(x)$.

This lack of symmetry can be easily cured by a slight modification of
the interaction in the ghost sector: Consider the interaction described by
$$\Theta^{(1)}(x)=-{i\over 2}f_{a,b,c}\{:\AMa\ANb\Fmnc:(x)+{1\over 2}
\AMa\ub\dsm\tuc:(x)\} \eqno (6.12)$$
This operator differs from $T^{(1)}(x)$ only by a pure divergence
term (i.e. a term of the form $\dM H^\mu$ [8]) and a ${\cal Q}$-boundary
(i.e. a term of the form $\{{\cal Q},K\}_+$ [3,23]).
It therefore remains gauge invariant [3,8]. It additionally
commutes with 3 basis charges in $\{Q\}$:
$$ [Q_g,\Theta^{(1)}(x)]_-=[\Gamma,\Theta^{(1)}(x)]_-=
[\Omega,\Theta^{(1)}(x)]_-=0 \eqno (6.13)$$
and is, therefore, $C_g$ invariant,too.
Only the central charge $N$ does not commute with $\Theta^{(1)}(x)$.
The transformations $T_0$ generated by the 3 charges in the last eq.
can be written as
$$T_0=T(A_0),\quad {\rm with}\,\,\, {\rm det}A_0=0 \eqno (6.14)$$
i.e. $\{T_0\}$ is a faithful representation of ${\rm SL}(2,{\cal C})$.

A symmetry of a gauge theory living in a Krein space should also
preserve the indefinite form (5.8), i.e. should be implemented by
pseudo-unitary transformation. This condition selects those
transformation $S_0$ in $\{T_0\}$ which are simultaneously in
$\{S_J\}$. They can be written as
$$S_0=T(M_0),\quad {\rm with}\,\,\, {\rm det}M_0=0\,\land MM^k=1 \eqno
(6.15)$$
The matrices $M_0$ form the real three-dimensional Lie group
${\rm SU}(1,1)$ and $\{S_0\}$ is a faithful representation of this
group.

$\Theta^{(1)}(x)$ is, due to its symmetry in ghost space,
pseudo-antihermitian with respect to both Krein operators
$I$ and $J_g$ discussed in the last chapter:
$$\left[\Theta^{(1)}(x)\right]^K=\left[\Theta^{(1)}(x)\right]^*
=-\Theta^{(1)}(x) \eqno (6.16)$$
In fact, it would be pseudo-antihermitian with respect to any Krein
operator which originates from $I$ by an arbitrary unitary
transformation in $\{T\}$.

Let us also state that the representation of $P_+^\uparrow$
discussed in the end of chapter 2 is pseudo-unitary: Fore, it is
unitary and it commutes with all charges $Q(a)$ (these are
$P_+^\uparrow$-scalars) and transformations $T(A)$. It is, therefore,
pseudo-unitary with respect to any Krein operator chosen in
$\{T\}$, too.

We close by remarking that the reasoning which leads to the
introduction of the Krein operator $J_A$ (5.9,5.10) in the
Yang-Mills sector of the theory is very different: The adjoint
Yang-Mills fields $\left[\Ama(x)\right]^+$ do have causal commutators there.
The representation of $P_+^\uparrow$, however, is not unitary, and
$J_A$ has to be introduced for the reason of covariance. This already
happens in QED [15].

\vskip 1.5cm

{\bigtype{\bf Acknowledgments - }} I would like to thank: Prof. L.
\'O Raifeartaigh for his kind interest in this work and for his
valuable comments; Dr.C. Ford for proofreading the manuscript;
the Swiss National Science Foundation for financial support;
and the Dublin Institute of Advanced Studies for warm hospitality.

\vskip 1.5cm
{\bigtype{\bf References}}
\vskip 0.2cm

\noindent [1] L.D. Faddeev, A.A. Slavnov: Gauge Fields, An Introduction To
Quantum Theory, second edition,

\noindent \hphantom{[1] }Addison-Wesley 1991

\noindent [2] C. Itzykson, J.B. Zuber: Quantum Field Theory, Mc Graw Hill 1980

\noindent [3] N. Nakanishi, I. Ojima: Covariant Operator Formalism Of Gauge
Theories And Quantum Gravity,

\noindent \hphantom{[3] }World Scientific 1990

\noindent [4] G. Velo, A.S. Wightman (eds.): Renormalization Theory, D.Reidel
1976

\noindent [5] F. Strocchi: Phys.Rev.D17, p.2010, 1978

\noindent [6] N.N. Bogolubov, A.A.Logunov, A.I. Oksak, I.I. Todorov:
General Principles Of Quantum Field Theory,

\noindent \hphantom{[6] }Kluwer 1990

\noindent [7] J. Bognar: Indefinite Inner Product Spaces, Springer 1974

\noindent [8] M. Duetsch, T. Hurth, F. Krahe, G. Scharf: Il Nuovo Cimento 106A
p.1029, 1993

\noindent [9] M. Duetsch, T. Hurth, F. Krahe, G. Scharf: Il Nuovo Cimento 107A
p.1029, 1994

\noindent [10] M. Duetsch, T. Hurth, G. Scharf: ZU-TH 29/94

\noindent [11] M. Duetsch, T. Hurth, G. Scharf: ZU-TH 35/94

\noindent [12] H. Epstein, V. Glaser: Annales de l'Institut Poincar\'e 29,
p.211, 1973

\noindent [13] N.N. Bogolubov, D.V. Shirkov: Quantum Fields, Benjamin-
Cummings 1984

\noindent [14] N.N. Bogolubov, D.V. Shirkov: Introduction To The Theory Of
Quantized Fields, Interscience 1959

\noindent [15] G. Scharf: Finite QED, Springer 1989

\noindent [16] M. Duetsch, F. Krahe, G. Scharf: Il Nuovo Cimento 103A, p.871,
1990

\noindent [17] S.S. Schweber: Relativistic Quantum Field Theory, Harper and
Row 1961

\noindent [18] F. A. Berezin: The Method Of Second Quantization, Academic
Press 1966

\noindent [19] K.O. Friedrichs: Mathematical Aspects Of The Quantum Theory Of
Fields, Interscience 1953

\noindent [20] F. Constantinescu: Distributionen und ihre Anwendung in der
Physik, Teubner 1974

\noindent [21] R.F. Streater, A.S. Wightman: PCT, Spin And Statistics, And All
That, Benjamin-Cummings 1978

\noindent [22] T. Hurth: Nonabelian Gauge Theories / The Causal
Approach, Diss. Uni Zuerich, 1994

\noindent [23] W.H. Greub: Linear Algebra, third edition, Springer 1967

\noindent [24] F. Strocchi, A.S. Wightman: J. Math. Phys. 15, p. 2198, 1974

\end